\documentclass{PoS}

\usepackage[varg]{txfonts}
\usepackage{multirow}
\usepackage{url}

\usepackage{graphicx}
\usepackage{subfig}

\usepackage{soul}

\newcommand{\FERMI}{\textit{Fermi}}

\newcommand{\Swift}{\textit{Swift}}
\newcommand{\Newton}{\textit{Newton}}
\newcommand{\NuSTAR}{\textit{NuSTAR}}
\newcommand{\RXTE}{\textit{RXTE}}

\title{Long term lightcurve of the BL Lac object 1ES\,0229+200 at TeV energies}

\ShortTitle{Long term TeV lightcurve of 1ES\,0229+200}

\author{\speaker{Gabriele Cologna}\,$^a$\,\thanks{Member of the International Max
    Planck Research School for Astronomy and Cosmic Physics at the University
    of Heidelberg (IMPRS-HD) and the Heidelberg Graduate School of Fundamental
    Physics (HGSFP).} , Mahmoud Mohamed\,$^a$\,\footnotemark[2] , Stefan J. Wagner\,$^a$\,,
    Alicja Wierzcholska\,$^a$\,\thanks{Mobility Plus Fellow} , Carlo Romoli\,$^b$\, for the
    H.E.S.S. Collaboration and Omar Kurtanidze\,$^{c,\,a}$\\ 
        \llap{$^{a}$} Landessternwarte, Universit\"at Heidelberg, K\"onigstuhl, D 69117 Heidelberg, Germany \\
        \llap{$^{b}$} Dublin Institute for Advanced Studies, 31 Fitzwilliam Place, Dublin 2, Ireland \\
        \llap{$^{c}$} Abastumani Observatory, Mt. Kanobili, 0301 Abastumani, Georgia\\
        E-mail: \email{gcologna@lsw.uni-heidelberg.de}}

\abstract{The high-frequency peaked BL Lac object 1ES\,0229+200 (z\,=\,0.14) was
  first detected in very high energy (VHE, E\,>\,100\,GeV) $\gamma$-rays by the
  H.E.S.S. (High Energy Stereoscopic System) collaboration in 2006
  \cite{hess_2007A&A...475L...9A}. No flux variability was reported
  in the initial study and its spectral characteristics have been used to
  derive constraints on the extragalactic background light (EBL,
  \cite{hess_2007A&A...475L...9A}) and on the intergalactic magnetic field
  (IGMF,
  e.g. \cite{Tavecchio_2010MNRAS.406L..70T,Dermer_2011ApJ...733L..21D,Taylor_2011A&A...529A.144T,Vovk_2012ApJ...747L..14V}). 
  1ES\,0229+200 has been observed with 
  H.E.S.S. for $\sim$130 hours from 2004 to 2013: the full dataset analysed
  with a more sensitive method will be presented here. The results indicate that the
  source is not constant and displays flux variability on yearly and monthly
  timescales. The existence of flux variability affects the derivation of the constraints on the IGMF.
  The H.E.S.S. observations cover several simultaneous multi-frequency campaigns
  and the VHE variations are compared with those reported in different bands.}

\FullConference{The 34th International Cosmic Ray Conference,\\
		30 July- 6 August, 2015\\
		The Hague, The Netherlands}

\begin{document}

\section{Introduction}

1ES\,0229+200 is a blazar, and a member of the high frequency peaked BL\,Lac objects
(HBL) subclass. It is hosted in an elliptical galaxy located at a redshift
$z$\,=\,0.1396. It was first discovered in X-rays by the \textit{Einstein} IPC
Slew Survey in 1992 \cite{EinsteinSurvey_1992ApJS...80..257E}. Its synchrotron
emission reaches 100\,keV and possibly up to
200\,keV\footnote{\url{http://bat.ifc.inaf.it/100m_bat_catalog/100m_bat_catalog_v0.0.htm}}. Despite
being relatively bright in optical and X-rays, it is faint in radio and
$\gamma$-rays. It has been detected at VHE above 500\,GeV for the first time
by H.E.S.S. in 2006 \cite{hess_2007A&A...475L...9A}, and only in 2011 at high
energies (HE, 100\,MeV\,<\,E\,<\,100\,GeV) between 1 and 300\,GeV, after more
than three years of \FERMI\, observations
\cite{Vovk_2012ApJ...747L..14V}. Contrary to what one could expect from its
blazar nature, 1ES\,0229+200 shows little variability at all energies, even if
it is known to have flux variations in the X-ray band of a factor of
$\sim$\,2. (e.g. \cite{Kaufmann_2011A&A...534A.130K}). Hints of variability at
TeV energies were reported on yearly and monthly timescales by the VERITAS
collaboration in 2014 \cite{Veritas_2014ApJ...782...13A}, while none was
detected in the source in the initial H.E.S.S. study.

1ES\,0229+200 has been one of the key sources for deriving constraints on 
the extragalactic background light (EBL, \cite{hess_2007A&A...475L...9A}) and on the 
intergalactic magnetic field (IGMF, e.g.
\cite{Tavecchio_2010MNRAS.406L..70T,Dermer_2011ApJ...733L..21D,Taylor_2011A&A...529A.144T,Vovk_2012ApJ...747L..14V})
thanks to the combination of its hard spectrum ($\Gamma_{PL}\sim$2.5,
\cite{hess_2007A&A...475L...9A,Veritas_2014ApJ...782...13A}) reaching
$\sim$\,10\,TeV and its considerable distance (for a TeV source).
The spectrum allows one to probe the near- and mid-infrared wavelengths of the EBL 
($\sim$\,2\,-\,20\,$\mu$m) at high optical depths ($\tau\sim$\,1\,-\,6 for $\gamma$-ray energies
between $\sim$\,1\,-\,10\,TeV). This is unique, since only a few other sources
have been observed at these energies. For the IGMF studies, a multi-TeV hard
spectrum in combination with high optical depths implies that a considerable
part of the emitted TeV photons will be absorbed via pair production through
their interaction with the EBL and re-emitted at (typically) GeV energies,
causing a surplus of HE emission. Lower limits on the strength of the IGMF can
be calculated measuring the GeV emission and making assumptions on the
intrinsic spectrum. Since reprocessing involves emission from an ensemble of
charged particles that are spread over an extended volume by the action of the
magnetic field, time delays are of the order of the light travel time of the
excess distance that the re-emitted photons must cover with respect to the
line of sight to the source. Time delays are also energy dependent, since a
given magnetic field will affect more strongly the trajectories of low-energy
particles. This holds true as long as the Compton cooling time is of the order
of the gyroradius. If it is too fast, re-emission is almost immediate and time
delay negligible. If it is too slow, this and not the magnetic field
determines the time delay. The observed reprocessed emission corresponds
therefore to the time-averaged original VHE emission. Constrains on the IGMF
hence rely on the assumption of a steady VHE flux on time scales of these
delays. Is the VHE flux variable, average values can be used as long as the
variability is shorter than the reprocessing time delay.

\section{Observations and Analysis}
\subsection{H.E.S.S.}
H.E.S.S. observations have been carried out almost yearly between 2004 and
2013, for a total of 354 observation runs\footnote{The term "run" refers to a
  single observation, with a typical exposure of 28 minutes.}.
 311 of them pass the standard
quality cuts for a total livetime of $\sim$\,133\,h (122\,h when corrected for
acceptance). The mean zenith angle is 45.2$^\circ$ and the mean offset from
the pointing position 0.51$^\circ$. The exposure was not homogeneous during
these observation years, ranging from 3.8\,h in 2004 to 47.1\,h in 2006. Refined quality
selection criteria allowed for a larger dataset for the years 2004-2006
than the one published in \cite{Aharonian_1st_ul_2005A&A...441..465A} and
\cite{hess_2007A&A...475L...9A}.
Part of the 2009 and 2013 data belong to multiwavelength (MWL) campaigns.
The first one was organized with XMM-\Newton\, and ATOM on August 21 and 23, 2009,
the second one with \NuSTAR, \Swift-XRT, MAGIC and VERITAS on October 1, 5 and 11, 2013.
Data reduction has been performed using the Model analysis\footnote{software version
  paris-0-8-24, DSTs Prod26} \cite{deNaurois2009APh....32..231D} with
\textit{Standard} cuts. For the background determination for the spectral and temporal analyses, 
the \textit{Reflected Region Background} method \cite{Berge2007A&A...466.1219B} was used.
The source is detected with a significance of 18.1 standard deviations ($\sigma$) for the whole dataset.
The results for the total dataset and for every year of observation can be found in
Tab.\ref{tab:counts_years}, as well as in Tab.\ref{tab:counts_2006_2009} for the 2006 and 2009
observing periods.
\begin{table}[tp]
  \centering
  \caption{Statistics of the yearly and total H.E.S.S. observations.}
  \label{tab:counts_years}
  \small
  \begin{tabular}{c|ccccccc}
    \hline
    \hline       \\[-0.8em]
    \multirow{2}{*}{Period} & Dates & Live Time & ON     & OFF    & excess & Significance & Flux (>\,580\,GeV)            \\  
                            & (MJD) & (h)       & counts & counts &        & ($\sigma$)   & (10$^{-13}$\,cm$^{-2}$s$^{-1}$)  \\   
    \hline
    \hline  \\[-0.8em]
    2004\,-\,2013 & 53259\,-\,56606 & 132.5 & 1839 & 14138 & 699 & 18.1   & 6.2\,$\pm$\,0.5   \\[0.2em]
    \hline  \\[-0.8em]
    2004 & 53259\,-\,53317 &  3.8     &  47 &  349  &  18    &  2.9       & 3.5\,$\pm$\,1.9    \\[0.2em] 
    2005 & 53613\,-\,53649 & 16.4     & 169 & 1518  &  42    &  3.4       & 2.5\,$\pm$\,1.0    \\[0.2em] 
    2006 & 53967\,-\,54088 & 47.1     & 750 & 4577  & 368    & 15.8       & 9.0\,$\pm$\,0.8    \\[0.2em] 
    2007 & 54322\,-\,54336 & 10.5     & 137 & 1080  &  47    &  4.4       & 5.6\,$\pm$\,1.7    \\[0.2em] 
    2008 & 54681\,-\,54789 & 12.2     & 149 & 1420  &  31    &  2.6       & 4.0\,$\pm$\,1.6    \\[0.2em]   
    2009 & 55063\,-\,55151 & 19.0     & 215 & 1825  &  63    &  4.6       & 5.2\,$\pm$\,1.4    \\[0.2em] 
    2011 & 55801\,-\,55909 &  7.4     & 112 & 918   &  35    &  3.6       & 1.9\,$\pm$\,2.0    \\[0.2em] 
    2013 & 56514\,-\,56606 & 17.8     & 241 & 2189  &  95    &  6.9       & 7.4\,$\pm$\,1.3    \\[0.2em] 
    \hline
    \hline
    \multicolumn{8}{l}{\parbox{1.0\textwidth}{\footnotesize{Note: the normalization factor $\beta$ between the ON and OFF area is 1/12 for all years with the exception of 2013 1/15.}}}   \\[0.2em] 
  \end{tabular}\\
\end{table}

\begin{table}[tp]
  \centering
  \caption{Statistics of the 2006 and 2009 observation periods.}
  \label{tab:counts_2006_2009}
  \small
  \begin{tabular}{l|ccccccc}
    \hline
    \hline       \\[-0.8em]
    \multirow{2}{*}{Period} & Dates & Live Time & ON     & OFF    & excess & Significance & Flux (>\,580\,GeV)            \\  
                            & (MJD) & (h)       & counts & counts &        & ($\sigma$)   & (10$^{-13}$\,cm$^{-2}$s$^{-1}$)  \\  
    \hline
    \hline  \\[-0.8em]
                            & \multicolumn{7}{c}{2006}               \\[0.2em]
    \cline{2-8}  \\[-0.8em]  
    August    & 53967\,-\,53977 & 13.5     & 226 & 1581  &  94    &  7.1       & 8.1\,$\pm$\,1.6       \\[0.2em]   
    September & 53994\,-\,54005 & 16.0     & 329 & 1813  & 178    & 11.9       & 12.2\,$\pm$\,1.6      \\[0.2em]   
    November  & 54048\,-\,54063 & 12.5     & 146 &  813  &  78    &  7.8       & 6.9\,$\pm$\,1.4       \\[0.2em]   
    December  & 54077\,-\,54088 &  5.1     &  49 &  370  &  18    &  2.9       & 6.8\,$\pm$\,2.0       \\[0.2em]   
    \hline
    \hline  \\[-0.8em]
                            & \multicolumn{7}{c}{2009}               \\[0.2em]
    \cline{2-8}  \\[-0.8em] 
    August    & 55063\,-\,55074 & 9.3      & 72  & 815   &  4     & 0.5        & -0.5\,$\pm$\,1.5        \\[0.2em]   
    October   & 55115\,-\,55121 & 5.1      & 73  & 561   & 26     & 3.4        & 7.0\,$\pm$\,2.9         \\[0.2em]   
    November  & 55145\,-\,55151 & 4.6      & 70  & 449   & 33     & 4.5        & 16.6\,$\pm$\,3.8        \\[0.2em]   
    \hline
    \hline \\[-0.8em] 
    \multicolumn{8}{l}{\parbox{1.0\textwidth}{\footnotesize{Note: the normalization factor $\beta$ between the ON and OFF area is 1/12 for all periods.
          The observing periods are defined by the lunar cycle and do not always strictly match the given month, which
          is given for an easier identification.}}}    \\[0.2em] 
  \end{tabular}\\
\end{table}

Spectral analyses have been carried out on different data subsets using the
forward-folding technique \cite{Piron_2001A&A...374..895P} and a simple
power-law (PL) spectral shape. No spectral variability is detected on a yearly
nor monthly timescale. The lightcurves are derived adjusting the normalization 
of a PL with index $\Gamma=2.9$
to the $\gamma$-ray excess in every time bin. Only
energies above a common threshold of 580\,GeV are considered. The monthly and
yearly lightcurves are shown in the top two panels of Fig.\,\ref{fig:lightcurves}
and discussed in Sec.\,\ref{sec:results}. All results have been cross checked with 
a different software and calibration chain.

\subsection{High Energies}
\label{subsec:fermi}

HE data have been collected by the Large Area Telescope (LAT) 
onboard the \FERMI\, satellite starting August 4, 2008. 
Data up to April 27, 2015 have been analysed with the ScienceTool 
software package version \verb|v9r33p0|. Only events belonging to the 'Source' class 
within 15$^\circ$ from the position of 1ES\,0229+200 were selected. Moreover, cuts 
on zenith angle (100$^\circ$), rocking angle (52$^\circ$) and distance from the Sun 
(5$^\circ$) were applied. For the binned maximum-likelihood spectral analysis, the 
instrument response functions \verb|P7REP_SOURCE_V15| were used, together with the 
publicly available standard Isotropic and Galactic diffuse emission background models 
iso\_source\_v05.txt and \verb|gll_iem_v05 rev1.fit|.

1ES\,0229+200 is detected with a significance of 10.5\,$\sigma$ in the energy
range 100\,MeV - 500\,GeV, compared to the 6.7\,$\sigma$ between 1 and
300\,GeV in \cite{Vovk_2012ApJ...747L..14V}. The low flux in the \FERMI-LAT
band does not permit detailed temporal studies on monthly
timescales. Nonetheless, the variability index\footnote{TS$_{\rm{var}}=2
    \Sigma_i[$log$\mathcal{L}_i(F_i)$\,-\,log$\mathcal{L}_i(F_{\rm{const}})]$
    where $F_{\rm{const}}$ and $F_i$ are the average source flux and the flux
    of the \textit{i}-th time bin. No limits on the bin significance are
    required.} for the monthly lightcurve has a value of 157 for 81 
degrees of freedom, which indicates variability at the 5\,$\sigma$ level.

\subsection{X-rays}
\label{subsec:swift}

In this work, X-ray data from \Swift, \RXTE\, and XMM-\Newton\, are
used\footnote{\NuSTAR\, data are not taken into account here since they are
  subject of a dedicated paper, in preparation.} (Fig.\,\ref{fig:lightcurves},
third and fourth panels). The 30-days bin \Swift-BAT lightcurve between 15 and
85\,keV is derived from the 15-days bin one in the 66 months Palermo BAT
Catalog\footnote{\url{http://bat.ifc.inaf.it/bat_catalog_web/66m_bat_catalog.html}}. The
2008 and 2009 XMM-\Newton\, and \Swift-XRT, as well as the 2010 \RXTE\, data
are taken from \cite{Kaufmann_2011A&A...534A.130K}, while the whole 2010-2011
\RXTE\, lightcurve is taken from \cite{RXTE_2013ApJ...772..114R}. Finally, the
most recent \Swift-XRT data have been collected between the nights of October
1-2 and 10-11, 2013 during the aforementioned MWL campaign.  

All the available \Swift-XRT data collected between 2008 and 2015 (ObsIDs
00031249001-00031249050 and 00080245001-00080245006)
were analysed using the HEASoft software package
v.\,6.16\footnote{\url{http://heasarc.gsfc.nasa.gov/docs/software/lheasoft}}
with CALDB v.\,20140120. All the events were cleaned and calibrated using the
\verb|xrtpipeline| task and the data in the 0.3-10\,keV energy range with
grades 0-2 for WT mode and 0-12 for PC mode were analysed. The lightcurve flux
points were calculated from the spectra of single snapshots integrating
between 2 and 10\,keV. These were derived in the following way: the data were
grouped using the \verb|grappha| tool to have a minimum of 30 counts/bin and
then fit using XSPEC v.\,12.8.2 with a single power-law model and Galactic
hydrogen absorption fixed to $n_H=8.06\times 10^{20}$\,cm$^{-2}$
\cite{Kalberla05_gal_abs}.

\subsection{Optical}
\label{subsec:optical}

Optical monitoring of 1ES\,0229+200 has been carried out with the ATOM
telescope \cite{ATOM_2004AN....325..659H} between 2007 and 2012. 
ATOM is a 75\,cm altazimuth telescope equipped with an Apogee Alta U47 camera
(Apogee Alta E47+ between 2004 and 2011) 
and works in robotic mode. It is part of
the H.E.S.S. project and is used mainly for optical monitoring of variable
$\gamma$-ray sources as a constant support for MWL observations.
The observations for 1ES\,0229+200 have been carried out typically once 
every three nights during the August-December visibility window. 
The standard reduction, the photometry and the source calibration 
are done with an automatic pipeline. The photometry 
uses an aperture of 4$^{''}$, while two reference stars in the vicinity of the source and 
present on the same frame are used for the calibration.
During the 2013 MWL campaign ATOM was not operational and data were collected with 
the 70\,cm telescope of the Abastunami Observatory (Georgia), equipped with an
Apogee 6E camera. Observations taken with a R Cousins filter 
have been analysed with the Daophot II reduction software using
an aperture diameter of 10$^{''}$.

\section{Results and discussion}
\label{sec:results}
\begin{figure}[tp]
  \centering
  \includegraphics[width=1.0\columnwidth]{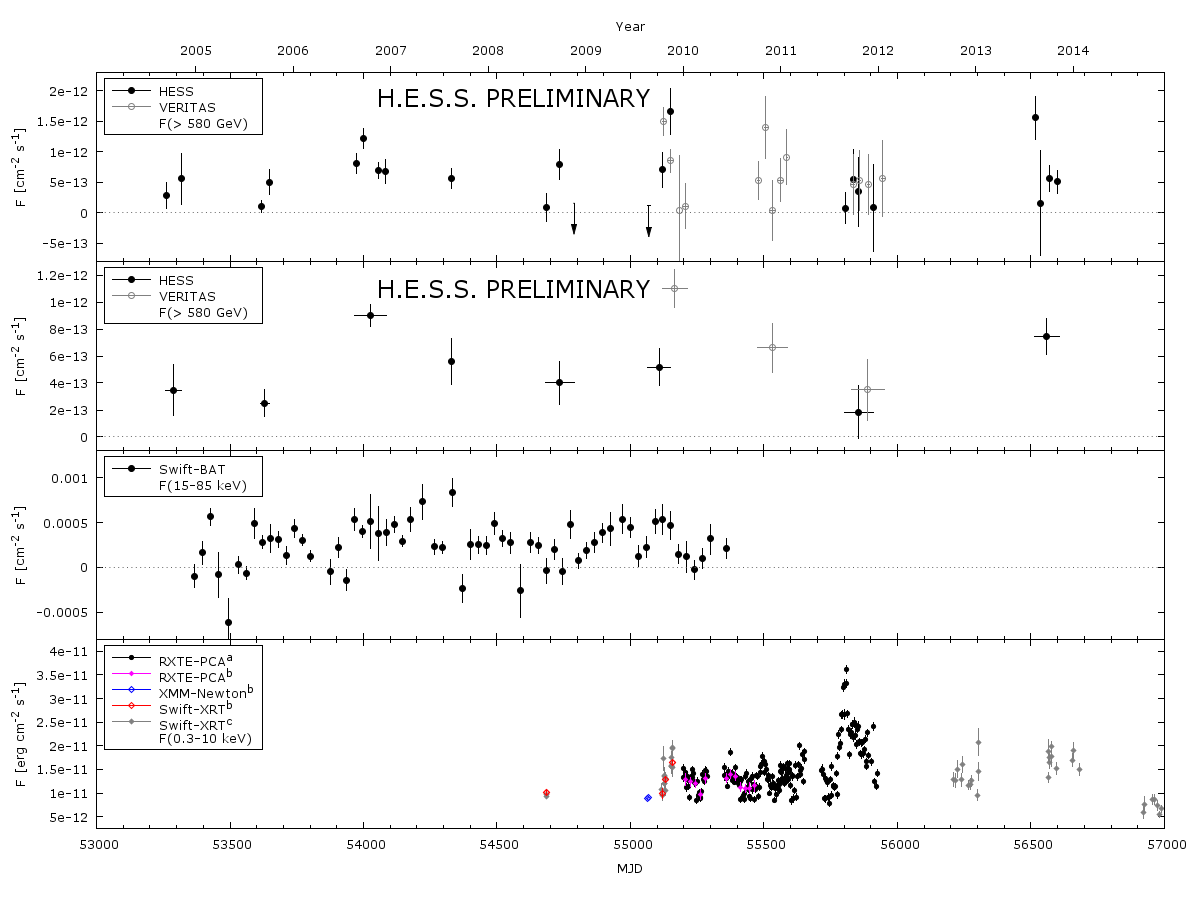}
  \caption{Lightcurves of the BL\,Lac object 1ES\,0229+200 in different energy
    bands between 2004 and 2013. From \textit{top} to \textit{bottom}: monthly and yearly
    H.E.S.S. lightcurves above 580\,GeV - the VERITAS values from \cite{Veritas_2014ApJ...782...13A}
    are also depicted as comparison; hard X-rays monthly lightcurve between 15 and 85\,keV
    from \Swift-BAT (the Palermo BAT Catalogue); soft X-ray lightcurve between
    2 and 10\,keV for different instruments: $^{a)}$ 2010-2012 \RXTE\, dataset from \cite{RXTE_2013ApJ...772..114R},
    $^{b)}$ XMM-\Newton, \Swift-XRT and \RXTE\, data from \cite{Kaufmann_2011A&A...534A.130K},
    $^{c)}$ \Swift-XRT data analysed in this work. The source is clearly
    variable in all energy bands. The discrepancy between the H.E.S.S. and VERITAS points
    in 2009 is explained by the non identical observation windows of the two instruments (see text).
    Three point at MJD 53831, 55331 and 55377 have been removed for clarity from the \Swift-BAT plot because of 
    their large negative fluxes or error bars.}
  \label{fig:lightcurves}
\end{figure}

The Model analysis is more sophisticated and sensitive than the Hillas
analysis \cite{Hillas_1985ICRC....3..445H} used in the previous
publications. Its use permits therefore to have a much higher sensitivity to
small flux variations. 
The monthly and yearly VHE lightcurves (top two panels of
Fig.\,\ref{fig:lightcurves}) show an emission which is not constant, neither
on a yearly, nor on a monthly timescale. A fit to a constant value yields a
$\chi^2$ of 84.3 (33.3) for 22 (7) degrees of freedom for the monthly (yearly)
lightcurve. This translates in a probability of 5$\times$10$^{-9}$ and
2$\times$10$^{-5}$ respectively. The values for the fractional variability
(\cite{Vaughan_2003MNRAS.345.1271V}, though whose errors are calculated as in
\cite{Poutanen_2008MNRAS.389.1427P}) are then 0.60\,$\pm$\,0.15 and
0.39\,$\pm$\,0.11. Variability is detected on a monthly timescale also
  between 0.58 to 1\,TeV and above 1\,TeV (probability of a constant flux of
  1$\times$10$^{-4}$ and 2$\times$10$^{-5}$, respectively).

In 2009, H.E.S.S. and VERITAS observed the source contemporaneously (though on
different nights). The time range spanned by the VERITAS observations was
significantly longer than the H.E.S.S. one. Taking the windows into account,
the measurements of both experiments are consistent with an increasing flux that
reaches its maximum just before MJD 55150 and then decreases. The comparison
of the VHE and X-ray data in 2009 \cite{Kaufmann_2011A&A...534A.130K} is very
interesting. The source is in very low state for both H.E.S.S. and
XMM-\Newton\, during the MWL campaign (MJD $\sim$55065). This changes in the
following three months, when \Swift-XRT detects a factor $\sim$2 flux increase
\cite{Kaufmann_2011A&A...534A.130K} (also seen by \Swift-BAT), which is
mirrored by an enhancement in the emission in $\gamma$-rays of a factor 2 or 3
(Fig.\,\ref{fig:lightcurves}). In all three wavebands, this state is followed
by a decrease in the emission. Despite the limited sensitivity of \Swift-BAT
in the hard X-ray band, it is interesting to notice that the three periods of
low VHE fluxes in 2008 and 2009 are mirrored also in this band. The R and B
bands optical lightcurves, instead, are constant within the errors and do not
show any variations. During the second MWL campaign in 2013 (MJD $\sim$56570),
1ES\,0229+200 seems to be relatively bright in X-rays and on its average value in TeV. 

A correlation in the X-ray-VHE emission is actually expected by the SSC model,
implying that the two components are generated by the same electron
population. On the other hand, the big peak shown by \RXTE\, in 2011 (MJD
$\sim$55800) does not have a clear correspondence in TeV. This could be
explained by the very low exposure in three out of the four periods of
H.E.S.S. observations, or by a period of high activity of one of the other 
X-ray sources present in the field of view of \RXTE\,
(e.g. 1RXS\,J023558.0+201215, 1RXS\,J023427.5+192247 or one of the numerous XMM
sources). A third possibility could be the presence of a
second X-ray emitting zone, not related to any $\gamma$-ray
emission. Finally, also Klein-Nishina suppression could play a role
  under certain conditions.

The confirmation of VHE variability on these timescales (already suggested in
\cite{Veritas_2014ApJ...782...13A}) affects the determination of constrains on
the IGMF. Focusing on the detected variability, if the $\sim$600\,GeV
emission consists of secondary photons ("maximal" case in
\cite{Taylor_2011A&A...529A.144T}), then a measurement of the IGMF can
actually be done. Extrapolating Fig.\,2 of \cite{Taylor_2011A&A...529A.144T}
to shorter time delays, one measures the IGMF to be B$_{\rm{IGMF}}\sim 3\times10^{-16}$\,G.

Although it has been stated that TeV $\gamma$-rays from 1ES\,0229+200 could be explained by 
proton cascade emission (e.g. \cite{essey_prot_casc_2011ApJ...731...51E}), the variability 
above 1 TeV implies that the radiation of this source is dominated by primary $\gamma$-rays.
Future studies investigating further the energy dependendent variability will 
allow the contribution of the cascade emission to the flux to be probed more 
deeply for both hadronic and leptonic models.

\section{Conclusion}
The 10 years long H.E.S.S. monitoring of the blazar 1ES\,0229+200 between 2004
and 2013 has been presented for the first time in a MWL context. Clear
variability is detected at VHE on monthly and yearly timescales. A hint
of correlation between TeV and X-ray emission comes from the contemporaneous
observations with XMM-\Newton\, and with \Swift\, in 2009: the fluxes increase
in all energy bands 
in similar way over four months. This supports an SSC emission model for the
X-ray and VHE emission in this source. The VHE monthly flux variability
affects the derivation of lower limits on the IGMF. Assuming the emission at
$\sim$600\,GeV to be of secondary photons, one can actually obtain a
measurement of the IGMF of B$_{\rm{IGMF}}\sim 3\times10^{-16}$\,G.

\section{Acknowledgments}
The support of the Namibian authorities and of the University of Namibia
in facilitating the construction and operation of H.E.S.S. is gratefully
acknowledged, as is the support by the German Ministry for Education and
Research (BMBF), the Max Planck Society, the German Research Foundation (DFG), 
the French Ministry for Research,
the CNRS-IN2P3 and the Astroparticle Interdisciplinary Programme of the
CNRS, the U.K. Science and Technology Facilities Council (STFC),
the IPNP of the Charles University, the Czech Science Foundation, the Polish 
Ministry of Science and  Higher Education, the South African Department of
Science and Technology and National Research Foundation, and by the
University of Namibia. We appreciate the excellent work of the technical
support staff in Berlin, Durham, Hamburg, Heidelberg, Palaiseau, Paris,
Saclay, and in Namibia in the construction and operation of the
equipment. 

O.M.K acknowledges financial support by the by Shota Rustaveli National
Science Foundation under contract FR/577/6-320/13.

This research has made
use of the Palermo BAT Catalogue and database operated at INAF\,-\,IASF
Palermo, and of the lightcurves provided by the University of California, San
Diego Center for Astrophysics and Space Sciences, X-ray Group
(R.E. Rothschild, A.G. Markowitz, E.S. Rivers, and B.A. McKim), obtained at
\url{http://cass.ucsd.edu/~rxteagn/}. 

\bibliographystyle{JHEP}
\bibliography{1es0229_reduced}

\providecommand{\href}[2]{#2}\begingroup\raggedright\begin{thebibliography}{10}

\bibitem{hess_2007A&A...475L...9A}
F.~{Aharonian et al.}, {\it {New constraints on the mid-IR EBL from the HESS
  discovery of VHE {$\gamma$}-rays from 1ES 0229+200}},  {\em \aap} {\bf 475}
  (2007) L9--L13, [\href{http://arxiv.org/abs/0709.4584}{{\tt
  arXiv:0709.4584}}].

\bibitem{Tavecchio_2010MNRAS.406L..70T}
F.~{Tavecchio et al.}, {\it {The intergalactic magnetic field constrained by
  Fermi/Large Area Telescope observations of the TeV blazar 1ES0229+200}},
  {\em \mnras} {\bf 406} (2010) L70--L74,
  [\href{http://arxiv.org/abs/1004.1329}{{\tt arXiv:1004.1329}}].

\bibitem{Dermer_2011ApJ...733L..21D}
C.~D. {Dermer et al.}, {\it {Time Delay of Cascade Radiation for TeV Blazars
  and the Measurement of the Intergalactic Magnetic Field}},  {\em \apjl} {\bf
  733} (2011) L21, [\href{http://arxiv.org/abs/1011.6660}{{\tt
  arXiv:1011.6660}}].

\bibitem{Taylor_2011A&A...529A.144T}
A.~M. {Taylor}, I.~{Vovk}, and A.~{Neronov}, {\it {Extragalactic magnetic
  fields constraints from simultaneous GeV-TeV observations of blazars}},  {\em
  \aap} {\bf 529} (May, 2011) A144, [\href{http://arxiv.org/abs/1101.0932}{{\tt
  arXiv:1101.0932}}].

\bibitem{Vovk_2012ApJ...747L..14V}
I.~{Vovk et al.}, {\it {Fermi/LAT Observations of 1ES 0229+200: Implications
  for Extragalactic Magnetic Fields and Background Light}},  {\em \apjl} {\bf
  747} (2012) L14, [\href{http://arxiv.org/abs/1112.2534}{{\tt
  arXiv:1112.2534}}].

\bibitem{EinsteinSurvey_1992ApJS...80..257E}
M.~{Elvis et al.}, {\it {The Einstein Slew Survey}},  {\em \apjs} {\bf 80}
  (1992) 257--303.

\bibitem{Kaufmann_2011A&A...534A.130K}
S.~{Kaufmann et al.}, {\it {1ES 0229+200: an extreme blazar with a very high
  minimum Lorentz factor}},  {\em \aap} {\bf 534} (2011) A130,
  [\href{http://arxiv.org/abs/1109.3628}{{\tt arXiv:1109.3628}}].

\bibitem{Veritas_2014ApJ...782...13A}
E.~{Aliu et al.}, {\it {A Three-year Multi-wavelength Study of the
  Very-high-energy {$\gamma$}-Ray Blazar 1ES 0229+200}},  {\em \apj} {\bf 782}
  (2014) 13, [\href{http://arxiv.org/abs/1312.6592}{{\tt arXiv:1312.6592}}].

\bibitem{Aharonian_1st_ul_2005A&A...441..465A}
F.~{Aharonian et al.}, {\it {Observations of selected AGN with HESS}},  {\em
  \aap} {\bf 441} (2005) 465--472,
  [\href{http://arxiv.org/abs/astro-ph/0507207}{{\tt astro-ph/0507207}}].

\bibitem{deNaurois2009APh....32..231D}
M.~{de Naurois} and L.~{Rolland}, {\it {A high performance likelihood
  reconstruction of {$\gamma$}-rays for imaging atmospheric Cherenkov
  telescopes}},  {\em Astroparticle Physics} {\bf 32} (2009) 231--252,
  [\href{http://arxiv.org/abs/0907.2610}{{\tt arXiv:0907.2610}}].

\bibitem{Berge2007A&A...466.1219B}
D.~{Berge}, S.~{Funk}, and J.~{Hinton}, {\it {Background modelling in
  very-high-energy {$\gamma$}-ray astronomy}},  {\em \aap} {\bf 466} (2007)
  1219--1229, [\href{http://arxiv.org/abs/astro-ph/0610959}{{\tt
  astro-ph/0610959}}].

\bibitem{Piron_2001A&A...374..895P}
F.~{Piron et al.}, {\it {Temporal and spectral gamma-ray properties of Mkn 421
  above 250 GeV from CAT observations between 1996 and 2000}},  {\em \aap} {\bf
  374} (2001) 895--906, [\href{http://arxiv.org/abs/astro-ph/0106196}{{\tt
  astro-ph/0106196}}].

\bibitem{RXTE_2013ApJ...772..114R}
E.~{Rivers}, A.~{Markowitz}, and R.~{Rothschild}, {\it {Full Spectral Survey of
  Active Galactic Nuclei in the Rossi X-ray Timing Explorer Archive}},  {\em
  \apj} {\bf 772} (2013) 114, [\href{http://arxiv.org/abs/1306.4376}{{\tt
  arXiv:1306.4376}}].

\bibitem{Kalberla05_gal_abs}
P.~M.~W. {Kalberla et al.}, {\it {The Leiden/Argentine/Bonn (LAB) Survey of
  Galactic HI. Final data release of the combined LDS and IAR surveys with
  improved stray-radiation corrections}},  {\em \aap} {\bf 440} (2005)
  775--782, [\href{http://arxiv.org/abs/astro-ph/0504140}{{\tt
  astro-ph/0504140}}].

\bibitem{ATOM_2004AN....325..659H}
M.~{Hauser et al.}, {\it {ATOM - an Automatic Telescope for Optical
  Monitoring}},  {\em Astronomische Nachrichten} {\bf 325} (2004) 659--659.

\bibitem{Hillas_1985ICRC....3..445H}
A.~M. {Hillas}, {\it {Cerenkov light images of EAS produced by primary gamma}},
   {\em International Cosmic Ray Conference} {\bf 3} (1985) 445--448.

\bibitem{Vaughan_2003MNRAS.345.1271V}
S.~{Vaughan et al.}, {\it {On characterizing the variability properties of
  X-ray light curves from active galaxies}},  {\em \mnras} {\bf 345} (2003)
  1271--1284, [\href{http://arxiv.org/abs/astro-ph/0307420}{{\tt
  astro-ph/0307420}}].

\bibitem{Poutanen_2008MNRAS.389.1427P}
J.~{Poutanen}, A.~A. {Zdziarski}, and A.~{Ibragimov}, {\it {Superorbital
  variability of X-ray and radio emission of Cyg X-1 - II. Dependence of the
  orbital modulation and spectral hardness on the superorbital phase}},  {\em
  \mnras} {\bf 389} (2008) 1427--1438,
  [\href{http://arxiv.org/abs/0802.1391}{{\tt arXiv:0802.1391}}].

\bibitem{essey_prot_casc_2011ApJ...731...51E}
W.~{Essey}, O.~{Kalashev}, A.~{Kusenko}, and J.~F. {Beacom}, {\it {Role of
  Line-of-sight Cosmic-ray Interactions in Forming the Spectra of Distant
  Blazars in TeV Gamma Rays and High-energy Neutrinos}},  {\em \apj} {\bf 731}
  (Apr., 2011) 51, [\href{http://arxiv.org/abs/1011.6340}{{\tt
  arXiv:1011.6340}}].

\end{thebibliography}\endgroup

\end{document}